\documentclass{article}
\usepackage[utf8]{inputenc}
\usepackage{graphicx}
\usepackage{amsmath}
\usepackage{amssymb}
\usepackage[T1]{fontenc}
\usepackage{authblk}
\usepackage{xcolor}
\usepackage{soul}

\usepackage{cite}

\renewcommand{\vec}[1]{\mathbf{#1}}
\newcommand{\real}[1]{\mathcal{R}\Big[#1\Big]}
\newcommand{\imag}[1]{\mathcal{I}\Big[#1\Big]}

\title{End-to-end~nanophotonic~inverse~design~for\\
imaging and polarimetry}
\author[1]{Zin~Lin\thanks{zinlin@mit.edu}}
\author[2]{Charles~Roques-Carmes}
\author[1]{Rapha{\"e}l~Pestourie}
\author[2,3]{Marin~Solja\v{c}i\'{c}}
\author[4,5]{Arka~Majumdar}
\author[1]{Steven~G.~Johnson}
\affil[1]{Department of Mathematics, Massachusetts Institute of Technology, Cambridge MA 02138, USA}
\affil[2]{Research Lab of Electronics, Massachusetts Institute of Technology, Cambridge MA 02138, USA}
\affil[3]{Department of Physics, Massachusetts Institute of Technology, Cambridge MA 02138, USA}
\affil[4]{Department of Electrical and Computer Engineering, University of Washington, Seattle WA 98195, USA}
\affil[5]{Department of Physics, University of Washington, Seattle WA 98195, USA}

\date{\today}

\begin{document}

\maketitle

\begin{abstract}
By co-designing a meta-optical front end in conjunction with an image-processing back end, we demonstrate noise sensitivity and compactness substantially superior to either an optics-only or a computation-only approach, illustrated by two examples: subwavelength imaging and reconstruction of the full polarization coherence matrices of multiple light sources. Our end-to-end inverse designs couple the solution of the full Maxwell equations---exploiting all aspects of wave physics arising in subwavelength scatterers---with inverse-scattering algorithms in a single large-scale optimization involving $\gtrsim 10^4$ degrees of freedom. The resulting structures scatter light in a way that is radically different from either a conventional lens or a random microstructure, and suppress the noise sensitivity of the inverse-scattering computation by several orders of magnitude.  Incorporating the full wave physics is especially crucial for detecting spectral and polarization information that is discarded by geometric optics and scalar diffraction theory.
\end{abstract}

\section{Introduction}

\begin{figure}
    \centering
    \includegraphics[scale=0.55]{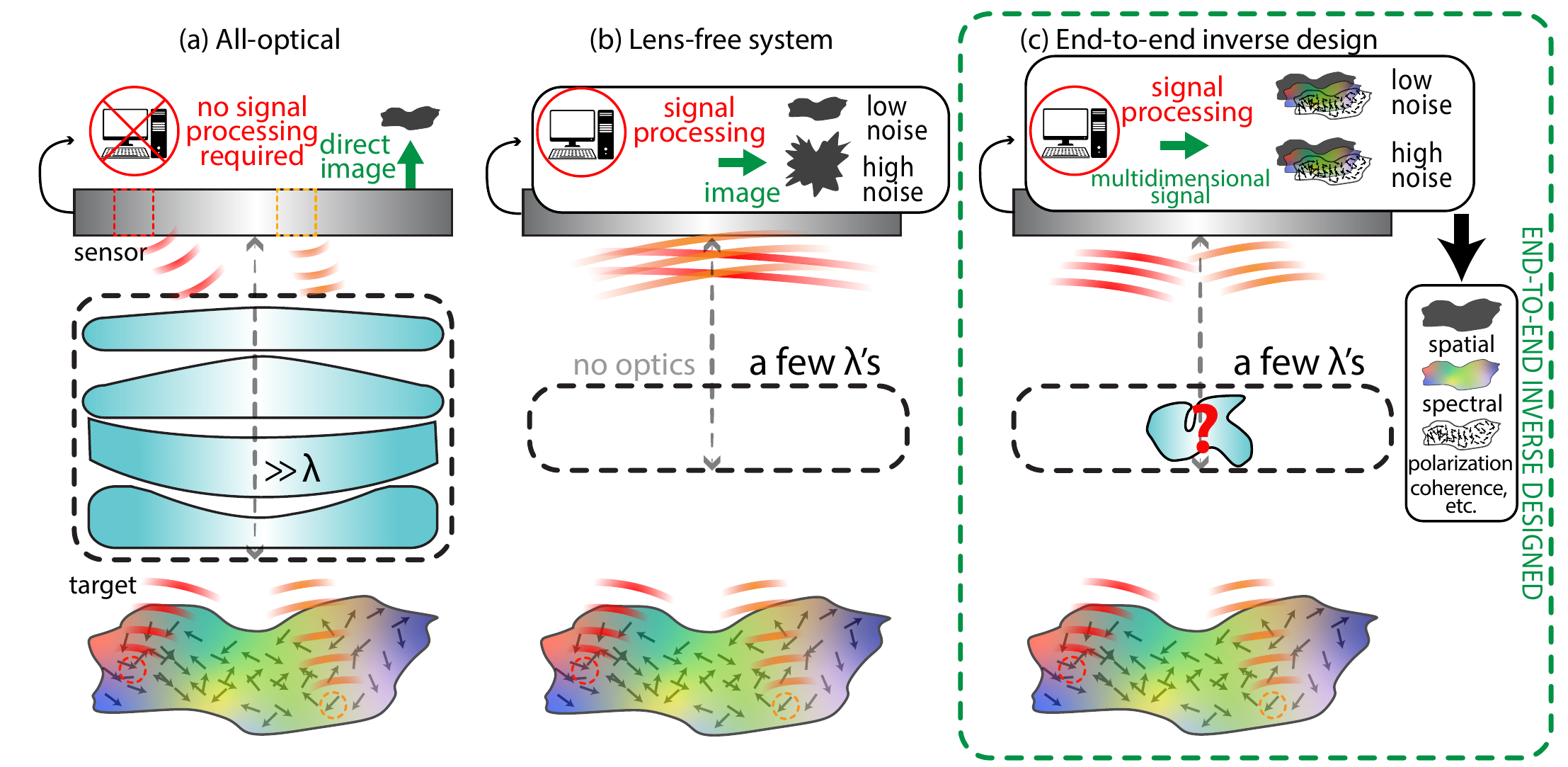}
    \caption{Comparison of three imaging modalities. (a) In traditional all-optical imaging, a bulky optical system focuses each point of the target on a different sensor pixel to directly produce a physical image; however, refractive or diffractive lenses are not designed for capturing the polarization or spectral content of the target. (b) In a compact lens-free system, the sensor directly records a blurry image while signal processing attempts to solve the resulting ill-posed (noise-sensitive) reconstruction problem; in this way, a spatial intensity profile of the target may be accurately reconstructed under sufficiently low noise conditions but polarization and spectral information cannot be retrieved. (c) In this work, we present an end-to-end inverse design approach, which optimizes a nanophotonic structure alongside the signal processing algorithm leading to an ultra-compact, noise-robust ``all-in-one'' system which may be used for not only imaging but also extracting polarization and spectral information.}
    \label{fig:schematic0}
\end{figure}

Computational imaging and computer vision plays an increasingly important role in modern technology, ranging from simplest image de-noising routines to state-of-the-art object recognition, robotic vision and machine intelligence algorithms with widespread demand in defense, medical as well as emerging Internet-of-things (IoT) industries. Traditional computer vision is exclusively driven by innovating the computational back end, and more recently, via deep learning and AI software. Little attention has been paid to the optical hardware at the front end beyond conventional lenses and diffraction gratings, in which light propagation is designed only by geometric optics~\cite{antipa2018diffusercam,sitzmann2018end,dun2020learned}. The full potential of wave physics has yet to be exploited for imaging device design in conjunction with computational reconstruction, especially for extracting spectral and polarization information that is mostly discarded by geometric optics. The last decade has seen explosive advances in understanding and manipulation of light waves and light-matter interactions at the most profound level of nano-materials, abetted by the development of efficient numerical modeling/design techniques as well as the advent of sophisticated nano-fabrication machinery. Those capabilities have been exploited for purely optical designs, such as metasurface lenses, that involve little or no computational post-processing~\cite{yu2014flat,khorasaninejad2016metalenses}. In this paper, we demonstrate the potential of 3D nanophotonics in the development of next-generation computer-vision technologies, in which conventional optics hardware is replaced by exquisitely designed nanophotonic structures; in particular, we propose to bring deeper and richer physics to computer vision by optimally tailoring a nanophotonic front end for a computational-imaging back end using a fully coupled inverse-design process, offering ultra-compact form factors as well as unprecedented capabilities for physical data acquisition and manipulation.

A conventional all-optical imaging system (Fig.~\ref{fig:schematic0}a) maps each point in a ``target'' space onto a separate sensor pixel, directly producing a faithful \emph{spatial} image but requires bulky optics, and also typically fails to capture detailed \emph{spectral} or \emph{polarization} content without additional filters. In another extreme, a compact lens-free system (Fig.~\ref{fig:schematic0}b) would directly detect a blurry image of the target and attempt to solve the subsequent ``inverse scattering'' problem (target reconstruction by, e.g., least square fitting), which is typically very ill-conditioned and hence sensitive to noise~\cite{tarantola2005inverse,gill2015computationally,asif2016flatcam,gill2013enabling,wang2009angle}. In this paper, we introduce an \emph{end-to-end} approach for inverse scattering (Fig.~\ref{fig:schematic0}c), in which a compact meta-optical structure is generated by large-scale inverse design of the full Maxwell equations \emph{coupled with} signal processing for target recovery, both for conventional spatial imaging and for spectral polarimetry.  First, we show that noise-tolerant subwavelength ($0.2\lambda$) far-field reconstruction of a collection of point sources is possible even with an ultra-compact ($2\lambda$-thick) imaging device.  Second, we demonstrate a ``multi-dimensional'' polarimeter that can resolve the full polarization states of multiple point sources at multiple frequencies. Specifically, we design meta-optical structures that generate well-conditioned (noise-robust) inverse-scattering problems, while exploiting a simple Tikhonov-regularization method (Sec.~\ref{sec:results}) to obtain subwavelength resolution without subwavelength focusing, or to enable multi-dimensional information extraction from a single-shot measurement. Accomplishing this requires that the optical ``inverse'' design problem, involving large-scale optimization over $\approx 10^4$ degrees of freedom, be coupled with the reconstruction algorithms (Sec.~\ref{sec:framework}).  That is, we perform ``end-to-end'' design in which the error $L(\varepsilon,p)$ of the reconstructed targets is jointly minimized as a function of both the microstructure ($\varepsilon$) and the reconstruction parameters ($p$).    Applying this approach to a two-dimensional (2D) example problem (Sec.~\ref{sec:results}), we obtain $0.22\lambda$ spatial resolution with a robust condition number (noise sensitivity) of only $\approx 10$, an improvement of $10^2$--$10^3$ over the condition numbers for lens-free or random (diffusing~\cite{antipa2018diffusercam}) scattering structures. Applying similar techniques to the polarimetry problem (Sec.~\ref{supercam}), we obtain a full-3D inverse-designed probe with a robust condition number of $\approx 6$ that can reconstruct nine-parameter polarization-coherence matrices of two point sources emitting at two frequencies.

Recent work in end-to-end computational imaging achieved improved image quality using regularized least-square image reconstruction in conjunction with scalar diffraction theory (rather than the full Maxwell equations) to design a phase plate (i.e., treated as locally uniform and neglecting multiple scattering)~\cite{sitzmann2018end}. 
Flat-optics meta-lenses~\cite{yu2014flat,khorasaninejad2016metalenses,chen2018broadband}, in contrast, have utilized more complete wave optics theory ranging from locally periodic~\cite{pestourie2018inverse,lin2019topology} or overlapping~\cite{lin2019overlapping} domain approximations to full Maxwell calculations~\cite{lin2018topology,chung2020high} coupled with optimization-based inverse design~\cite{lalau2013adjoint,jensen2011topology,molesky2018inverse}, exploiting local resonances and multiple scattering to achieve diffraction-limited focusing~\cite{bayati2020inverse,phan2019high}. Four-parameter Stokes imaging has also been demonstrated using a meta-optics polarization sorter in combination with a refractive lens~\cite{rubin2019matrix}. However, these works specified the focal point and/or the desired wavefront {\it a priori}, even with more complex focal patterns chosen to facilitate subsequent computational processing~\cite{guo2019compact,colburn2019simultaneous,colburn2020metasurface}, rather than performing a fully coupled end-to-end design.  There is also a vast body of work on computational image reconstruction~\cite{kino1987acoustic,greenbaum2012imaging}, but decoupled from the lens design (taking the optics as an immutable input rather than as design parameters).  In contrast, we couple the full Maxwell equations with the post-processing reconstruction \emph{during} the design process (Sec.~\ref{sec:framework}), so that an optimal wavefront is determined for each source to maximize reconstruction accuracy. Specifically, we demonstrate imaging with sub-wavelength resolution and multi-dimensional information extraction in ultra-compact form factors, a feat not possible using previously reported end-to-end computational imaging. In order to perform this optimization, we employ standard adjoint techniques from photonic inverse design~\cite{lalau2013adjoint,jensen2011topology,molesky2018inverse} combined with automatic-differentiation tools~\cite{maclaurin2015autograd} to obtain the sensitivity to changes in structural parameters $\varepsilon$ and reconstruction parameters $p$.   

\section{End-to-end framework}
\label{sec:framework}

\begin{figure}
    \centering
    \includegraphics[scale=1]{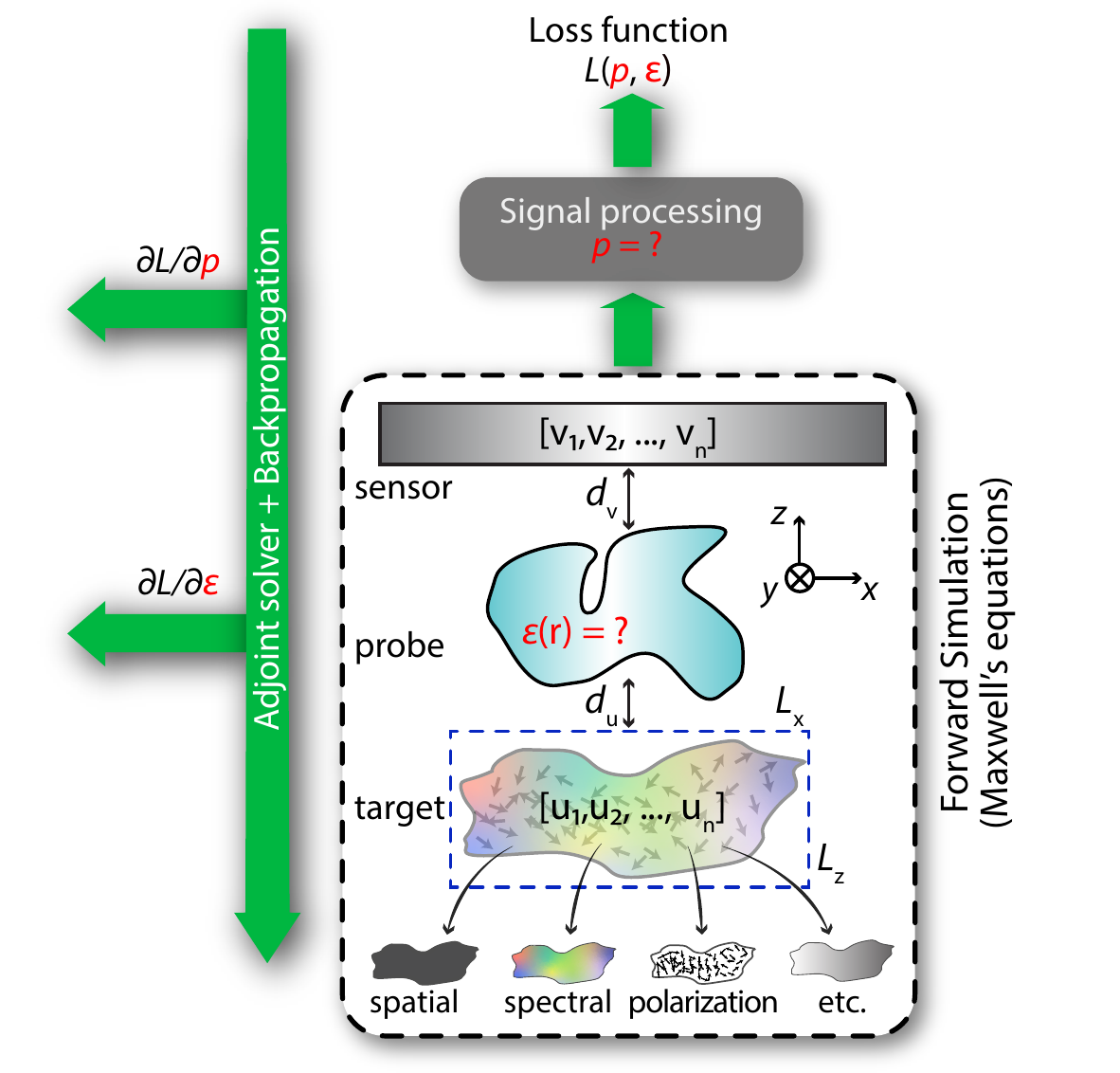}
    \caption{A schematic of the end-to-end inverse design framework. The target region of interest is characterized by an intensity vector $u$ of length $n$, containing spatial, spectral and/or polarization information. The photonic probe has a dielectric profile $\varepsilon(r)$ (to be determined via inverse design). The sensor, with $m$ pixels, records the raw image $v$. $u$ and $v$ are related by the forward scattering model: $v=G(\varepsilon)u+\eta$, where $G$ is a $m \times n$ matrix whose columns are extracted from the solution of the full Maxwell equations, and $\eta$ is a noise vector (e.g. sensor noise). $v$ is then fed into a signal processing algorithm parametrized by a vector $p$; the overall performance is evaluated by a loss function $L$ (e.g. mean square deviation from the ground truth). The processing may involve any operations including matrix-vector multiplications, nonlinear kernels, integro-differential equations or artificial neural networks; in particular, we consider the inverse scattering problem of estimating $u$ through regularized least-square minimization. End-to-end inverse design seeks optimal $\varepsilon$ and $p$ that optimizes the entire work-flow including both the forward model and the inverse problem; the gradients are obtained by backpropagation and adjoint methods.
    }
    \label{fig:schematic}
\end{figure}

Fig.~\ref{fig:schematic} shows a schematic of our proposed framework which can be applied to any wave-scattering problem including imaging, spectroscopy, polarimetry or any combination thereof. Here, the goal is to reconstruct a target $u$ in a pre-selected region of interest by computationally analyzing the captured image $v$ on a sensor. In between the sensor and the target region, we place a scattering structure, aka a \emph{photonic probe},  $\varepsilon(r)$ to be designed, at a ``working'' distance $d_u$ from the target and an ``image'' distance $d_v$ from the sensor. The state of the target is specified by a vector $u = [ u_1, ..., u_n ]$ containing spatial, spectral and/or polarization information, the details of which depend on the specific problem at hand.  The sensor has $m$ pixels with corresponding intensities (\emph{raw} image) $v=[v_1,...,v_m]$ given by the forward scattering model $v = G(\varepsilon) u + \eta$ where $G$ represents the solution of the Maxwell equations and $\eta$ is an \emph{additive} noise vector. For simplicity, we will consider zero-mean half-Gaussian white noise with non-zero standard deviation ($\eta \sim |{\cal N}(0,\sigma)|$; note that intensity noise $\eta$ is non-negative)~\cite{sitzmann2018end}, although our method can be easily adapted to other noise models (such as Poisson/shot noise) by calibrating the camera beforehand. We consider a planar sensor, which is the most common configuration in imaging, but our framework can readily be extended to arbitrary sensor topologies. The linear kernel $G$ is a $m \times n$ matrix whose columns are essentially point spread functions (PSF)~\cite{goodman2005introduction,antipa2018diffusercam} computed from the underlying Maxwell equations given a structure $\varepsilon(r)$.

The raw image $v$ is fed into a signal-processing algorithm to approximately reconstruct $u$~\cite{tarantola2005inverse}, in our case by a regularized least-square fit. That is, we find $\hat{u}$ such that $\hat{u} = \text{arg}~\text{min}_{\mu}~\Vert G\mu - v\Vert^2 + R(\mu) $. Here, $R$ is a regularization operator which serves to \emph{condition} a typically ill-posed inverse problem; essentially, $R$ incorporates any prior assumptions about $u$ (such as smoothness or sparsity) which ensure that the inverse problem has a stable unique solution~\cite{tarantola2005inverse}. In particular, we choose a Tikhonov ($L_2$) regularization $R(\mu) = \alpha \Vert \mu \Vert^2$ where $\alpha > 0$ is a regularization parameter to be determined, and $\hat{u}$ has a closed-form solution $\hat{u} = \left( G^TG + \alpha I\right)^{-1}G^Tv$~\cite{tarantola2005inverse}.   The noise sensitivity of the reconstructed $\hat{u}$ is characterized by the \emph{condition number} $\kappa(G)$ of the matrix $G$, which is a dimensionless quantity $\ge 1$ that is roughly proportional to the ratio of the $\Vert \hat{u} - u \Vert / \Vert u \Vert$ relative error to the input noise $\Vert \eta \Vert /  \Vert v \Vert$~\cite{trefethen1997numerical}. ($\kappa(G)$ can be computed as the ratio of the largest to smallest singular values of $G$.)
Many other variations are possible, such as $L_1$ ``sparse'' reconstruction~\cite{boyd2004convex} or artificial neural networks~\cite{szegedy2013deep,milletari2016v}. As we discuss in Sec.~\ref{sec:summary}, our approach extends easily to such techniques, even if the reconstruction problem does not have a closed-form solution or it involves a vast number of free parameters to be determined.   In general, a reconstruction algorithm is charactized by a vector $p$ of $P$ parameters; in this example, $p = [\alpha]$ and $P=1$. 

The end-to-end inverse design seeks optimal choices of $\varepsilon$ and $p$, which are tightly coupled by the end-to-end work-flow, in order to minimize the difference between the reconstructed $\hat{u}$ against the ground truth $u$. Specifically, we define a loss function $L(\varepsilon,p)$, here a mean-square error (MSE), such that $L = \langle \Vert u - \hat{u}\Vert^2 \rangle_{u,\eta}$ where $\langle \cdots \rangle_{u,\eta}$ denotes averaging (expected value) over many realizations of $u$ and $\eta$. The formulation can be now written as:
\begin{align}
    \text{min}_{\varepsilon,p} \quad L&=\langle \Vert u - \hat{u}\Vert^2 \rangle_{u,\eta} \\
    \hat{u} &= \left( G^TG + \alpha I\right)^{-1}G^Tv \\
    v &= G(\varepsilon) u + \eta
\end{align}
Here, the PSF matrix $G$ is extracted from the numerical solution of the Maxwell equations by any method.

In this paper, we consider the frequency-domain Maxwell equations with time-harmonic sources $e^{-i \omega t}$~\cite{jackson1999classical}:
\begin{align}
    \nabla \times \nabla \times E - \omega^2 \varepsilon(r) E = i \omega J.
\end{align}
solved by a finite-difference frequency-domain (FDFD) method~\cite{jin2015finite}.
For each voxel in the target region, $J$ is chosen as a point-source situated at the center of the voxel and the corresponding PSF is obtained by simulating the integrated electric field intensities $|E|^2$ over the sensor plane. The optimization over $\varepsilon$, $p$ require their gradients ${\partial L \over \partial \varepsilon}$, 
${\partial L \over \partial p}$, which can be found by back-propagation through the signal processing stage~\cite{bishop2006pattern} and adjoint sensitivity analysis~\cite{jensen2011topology,molesky2018inverse} of the Maxwell equations. We numerically implement these gradients by coupling an open-source automatic-differentiation packages~\cite{maclaurin2015autograd} with our own Maxwell adjoint solvers~\cite{lin2019topology}. 

\section{Imaging at sub-wavelength resolutions}
\label{sec:results}

\begin{figure}
    \centering
    \includegraphics[scale=0.75]{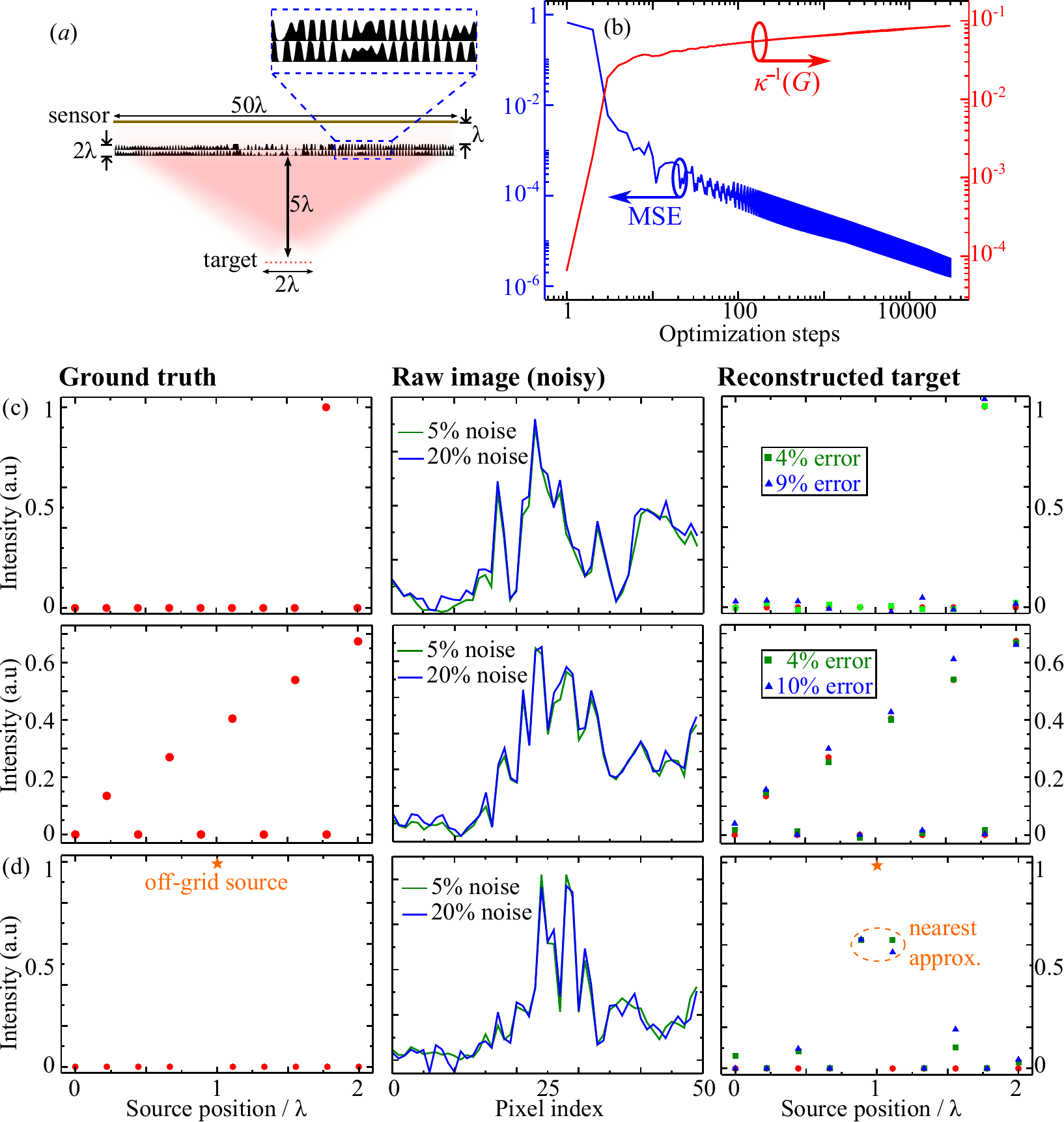}
    \caption{(a) Topology-optimized double-layer photonic probe ($\varepsilon\approx 2.3$) used to resolve a target of 10 point sources (Not drawn to scale). The probe is 50$\lambda$ wide and $\lambda$ thick, and is made up of freeform variable-height geometry. Note the scale bar. (b) Mean square error (MSE, blue line) and inverse condition number $\kappa^{-1}$ (red line) of the PSF matrix $G$, where $G$ is a $m \times n$ matrix with $m=50,~n=10$. $\kappa^{-1}$ steadily increases to around $0.08~(\kappa \approx 12)$, showing that the reconstruction becomes robust against noise. (c) Two example ground-truth targets are reconstructed under two different noise levels. A Gaussian noise $ \eta \sim \mathcal{N}(0,\sigma)$ is added to the image: $v = G u + \eta$. The standard deviation $\sigma$ is chosen as a percentage relative to the average PSF intensity $\bar{G}$; for example 5\% noise indicates $\sigma = 0.05 \bar{G}$. (d) The third example represents an \emph{off-grid} point source situated halfway between two grid nodes. The reconstruction intuitively produces a ``nearest-neighbors'' approximation. }
    \label{fig:highres}
\end{figure}

To demonstrate the capability of our framework, we consider an imaging problem at sub-wavelength resolutions. We consider a 2D problem $\varepsilon(x,z)$ (Fig.~\ref{fig:highres}) with $y$-polarized electric fields, so that the Maxwell equations are reduced to a scalar 2D Helmholtz equation. Specifically, we set $d_u=5\lambda$ (compact, but in the far field) and $d_v=\lambda$ (where near-field effects may be relevant) (Fig.~\ref{fig:highres}a) where $\lambda$ is the operating wavelength. Also, we discretize a 1D $2\lambda$-wide target region into $n=10$ point sources on an equi-spaced grid. Here, we assume incoherent illumination of the target region (as is common for imaging) so that only intensities need to be considered~\cite{goodman2005introduction}. Therefore, an arbitrary target residing within the region is described by an intensity vector $u = [ u_1, ..., u_n ]$ with a spatial resolution of $\Delta x_u = 0.222 \lambda$. (For targets at ``infinity,'' such as a photographic scene, the region of interest is an angular field of view and one can consider plane-wave sources instead of point sources.) The probe and sensor have a width of $50\lambda$ and the sensor contains $m=50$ pixels with a pixel size $\Delta x_v = \lambda$.  Although we have chosen these parameters for ease of demonstration, we note that this scenario is realizable using selective illumination~\cite{keller2008reconstruction}, slit apertures, a high-speed scanning mode, and line sensors~\cite{wang2018high} to produce 2D or even 3D images over a wide field of view.   More importantly, this system illustrates the essential ingredients of many important applications as discussed in Sec.~\ref{sec:summary}.

Although we have set $m>n$ (a nominally ``over-determined'' inverse problem), it is important to note that not any $\varepsilon(r)$ will lead to a well-conditioned (noise-robust) PSF matrix $G$. It is ill-advised to use a randomly-chosen $\varepsilon$ profile and directly invert $G$ because not every probe can resolve two point sources separated by a distance of $0.22\lambda$ and project measurably-distinct noise-tolerant PSFs onto a coarse-resolution sensor ($\Delta x_v \gg \Delta x_u$) one wavelength away from the probe (small $d_v$ leaves little room for conventional magnification). For example, we checked that a uniform $\varepsilon$ leads to $G$ with a condition number $\kappa(G) \approx 10^4$; even a disordered $\varepsilon$ with rapidly-varying fine features yields $\kappa(G) \approx 1000$. Both of these values represent orders of magnitude amplification of input noise in the output reconstruction, indicating that radical re-design of $\varepsilon$ is required.   

We show that our end-to-end framework can discover novel geometries $\varepsilon(r)$ with greatly reduced $\kappa(G)$, thereby rendering the inverse problem robust against noise. Here, the $\varepsilon$ degrees of freedom are a set of freeform variable heights~\cite{udupa2019voxelized} at each pixel within a double-layer design region made up of a low-permitivitty polymer material $\varepsilon_\text{polymer} \approx 2.3$ in air~(see Fig.~\ref{fig:highres}a). We have chosen these material settings because of rapidly maturing nano-scale 3D-printing technologies~\cite{buckmann2012tailored,zhan2019controlling} that would allow for exploration of such complex 3D geometries and are particularly suited for taking advantage of the full power of freeform topology optimization~\cite{molesky2018inverse,lin2019overlapping}. 

We employ stochastic gradient descent~\cite{bottou2010large} for optimizing $\varepsilon$ and $\alpha$ over $\gtrsim 10^4$ iterations including random noise $\eta$; we found that $\alpha$ stays close to an initial choice of $0.5$ while $\varepsilon$ evolves considerably during the course of optimization. In practice, we found that it works just as well to fix $\alpha$ with zero noise ($\eta=0$) as to vary $\alpha$ under many realizations of $\eta$ (note that $\alpha$ is closely related to the noise variance $\sigma^2$~\cite{o1986statistical}). Fig.~\ref{fig:highres}a exhibits a double-layer optimized design in a 3D-printable polymer-matrix (for example, Nanoscribe IP-DIP~\cite{gissibl2017refractive,fullager2017infrared}); each layer has thickness $\lambda/2$ and the minimum feature size is $\approx 0.04\lambda$, which may be challenging to fabricate at visible wavelengths but is feasible at longer wavelengths such as mid-wave and far-wave infra-red or even millimeter waves~\cite{yu2012potential,camayd2020multifunctional}. Fig.~\ref{fig:highres}b demonstrates that optimization rapidly improves both MSE ($\approx 10^{-6}$) and $\kappa(G) \approx 10$. Fig.~\ref{fig:highres}c shows two example ground truth targets being reconstructed under different noise levels. A Gaussian noise $ \eta \sim \mathcal{N}(0,\sigma)$ is added to the image: $v = G u + \eta$, where the standard deviation $\sigma$ is chosen as a percentage relative to the average PSF intensity $\bar{G}$; for example 5\% noise indicates $\sigma = 0.05 \bar{G}$. The low condition number ensures that the reconstruction errors are not amplified, remaining at $\approx 5\%$ and $10\%$ respectively for different $\sigma$'s.    

Like almost all computational imaging~\cite{donoho2006compressed}, this device is designed to reconstruct targets situated at a fixed grid (as in any camera with discrete pixels), but the accuracy degrades gracefully
for sources deviating from this grid. As shown in Fig.~\ref{fig:highres}d, even for the worst case of a light source that lies halfway between two grid points, the reconstructed image mostly divides the intensity between the two closest points.  (The error intensity at further points could be reduced if an L1 ``sparsifying'' reconstruction algorithm~\cite{donoho2006compressed} were used instead of L2 minimization.)   This degradation is known as ``gridding error'' in the computational-imaging community~\cite{strohmer2012measure}. A number of supplementary algorithms have been proposed to further improve the reconstruction for off-grid sources, including atomic-norm minimization~\cite{tang2013compressed} and coherence-inhibition schemes~\cite{fannjiang2012coherence}, which could be incorporated into end-to-end optical design if desired. Moreover, there are super-resolution imaging techniques such as photo-activation localization microscopy (PALM)~\cite{mockl2014super}, in which differential activation of sparse subsets of fluorescent molecules enables target recovery with arbitrary resolutions.  Yet another way to improve the differential target activation/reconstruction scenario would be to minimize the error over a collection of offset grids (instead of a single fixed grid of training data). 

Our results suggest that a low-index photonic micro-structure with a highly complex geometry can faithfully reconstruct an image down to deeply sub-wavelength resolutions (albeit over a \emph{finite} array of \emph{equi-spaced calibrated} point sources), while maintaining a sufficiently high signal-to-noise ratio. From a fundamental-physics perspective, we note that even though the probe is close to the target, the former is clearly not in the near field of the latter (since $d_u > \lambda/2$), which means evanescent fields from the target have negligible amplitude at the probe. Instead, the sub-wavelength resolution is made possible by the ability of the computational probe to distinguish the subtle \emph{differences} in spatial frequency components coming from adjacent point sources~\cite{harris1964resolving,harris1964diffraction,chen2020algorithmic}. Therefore, our approach is unlike negative-index metamaterial superlenses~\cite{pendry2000negative,zhang2008superlenses} or super-oscillatory lenses~\cite{huang2007optical}, which seek \emph{perfect point-to-point physical image formation} via amplification of evanescent waves or sub-diffraction-limit focal spots without the aid of computational reconstruction.


\section{Spatial + spectral + polarization extraction}
\label{supercam}

\begin{figure}
    \centering
    \includegraphics[scale=0.25]{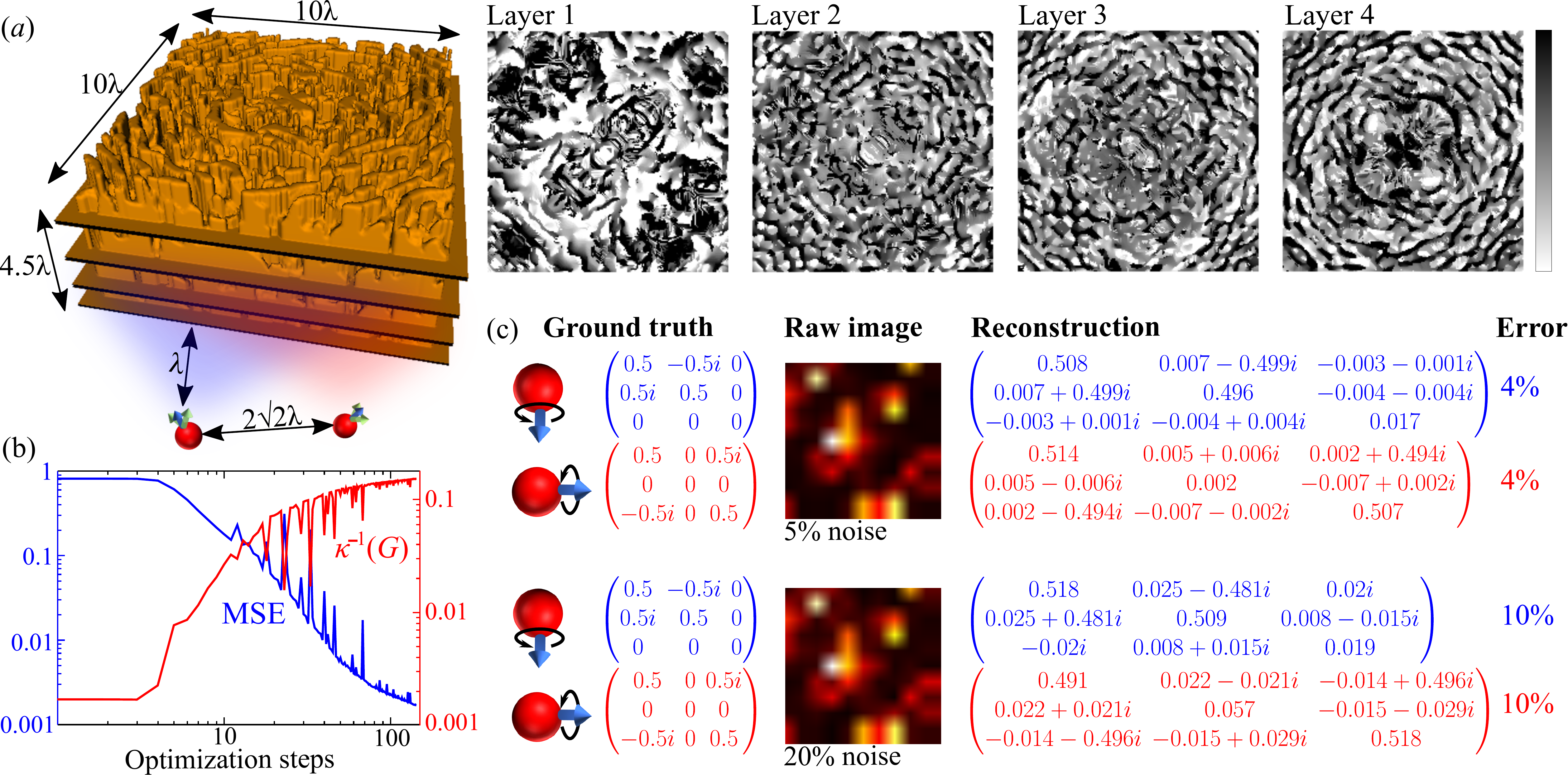}
    \caption{(a) Topology-optimized nanophotonic ``all-in-one'' probe for extracting spatial, spectral and polarization information. The probe can reconstruct the polarization coherence states of up to two point dipoles emitting at up to two spectral lines $\lambda$ and $1.1\lambda$. The dipoles must be positioned one $\lambda$ away from the probe and diagonally separated by $2\sqrt{2}\lambda$ away from each other whereas the sensor is located one $\lambda$ away on the other side of the probe. The probe consists of 4 layers of variable thickness polymer, whose gray-scale thickness profiles are also shown. (b) Mean square error (MSE, blue line) and inverse condition number $\kappa^{-1}$ (red line) of the PSF kernel $G$, where $G$ is a $m \times n$ matrix with $m=100,~n=36$. $\kappa_\text{opt}$ is found to be $\approx 6.4$. (c) As an example, two dipoles with two different polarization states and emitting at two wavelengths $\lambda$ (blue) and $1.1\lambda$ (red) are reconstructed under different noise levels. 
    }
    \label{fig:supercam}
\end{figure}

The ability to intimately manipulate the polarization states of light is a  hallmark of vectorial Maxwell photonics~\cite{arbabi2015dielectric,mueller2017metasurface}, which sets it apart from traditional geometric or diffractive optics. For example, in super-resolution microscopy with traditional lenses, the unresolved polarization state of a fluorescent molecule may affect localization accuracy and degrade the image recovery process, posing a nuisance in many imaging systems~\cite{backlund2014role,backlund2016removing}. Here, we show that end-to-end optimization can be used to design a nano-photonic polarimeter that can resolve the polarization coherence state of a fluorescent molecule, approaching theoretical upper bounds~\cite{tyo2002design}. In particular, the instantaneous polarization state of a point-dipole source (e.g., a fluorescent molecule or a solid-state quantum emitter, such as a quantum dot or color center) is specified by the complex-valued 3-element polarization vector $\mathbf{J} = [J_x,J_y,J_z]$. However, only the time-averaged intensities can be detected at optical frequencies, and the detectable polarization state of the dipole is described by a $3 \times 3$ coherence matrix~\cite{damask2004polarization} (equivalent to a matrix~\cite{scully1999quantum}):
\begin{align}
    \mathbf{D} = \langle \mathbf{J J^\dagger} \rangle =
    \begin{pmatrix}
    \langle |J_x|^2 \rangle & \langle J_x J_y^* \rangle & \langle J_x J_z^* \rangle \\
    \langle J_y J_x^* \rangle & \langle |J_y|^2 \rangle & \langle J_y J_z^* \rangle \\
    \langle J_z J_x^* \rangle & \langle J_z J_y^* \rangle & \langle |J_z|^2 \rangle
    \end{pmatrix}.
\end{align}
Here, $\langle \cdot \rangle$ denotes a time average. These nine coherence parameters associated are a natural generalization of the familiar four Stokes parameters~\cite{damask2004polarization} that characterize the polarization coherence state of a plane wave. 

The electric-field response $\vec{E}(\vec{r})$ of a nanophotonic structure $\varepsilon(\vec{r})$ in the presence of an arbitrarily polarized point dipole can be completely specified by the three ``basis'' fields, $\vec{u}_1$, $\vec{u}_2$ and $\vec{u}_3$, derived from $x$, $y$ and $z$-polarized test sources at the same location as the dipole: 
$\vec{E}(\vec{r},\varepsilon) = 
      J_x \vec{u}_1(\vec{r},\varepsilon) 
    + J_y \vec{u}_2(\vec{r},\varepsilon) 
    + J_z \vec{u}_3(\vec{r},\varepsilon) 
$. The integrated time-averaged electric-field intensity at the $i^\text{th}$ pixel on the sensor is then given by:
\begin{align}
    \int_i~\langle \big|\vec{E}\big|^2 \rangle 
    &= \int_i~\Bigg\{ 
    |\vec{u}_1|^2~\langle|J_x|^2\rangle + |\vec{u}_2|^2~\langle|J_y|^2\rangle + |\vec{u}_3|^2~\langle|J_z|^2\rangle \\
    & + 2\real{\vec{u}_1\cdot\vec{u}_2^*} \left( \real{\langle J_x J_y^* \rangle} \right) 
      - 2\imag{\vec{u}_1\cdot\vec{u}_2^*} \left( \imag{\langle J_x J_y^* \rangle} \right) \\
    & + 2\real{\vec{u}_1\cdot\vec{u}_3^*} \left( \real{\langle J_x J_z^* \rangle} \right) 
      - 2\imag{\vec{u}_1\cdot\vec{u}_3^*} \left( \imag{\langle J_x J_z^* \rangle} \right) \\
    & + 2\real{\vec{u}_2\cdot\vec{u}_3^*} \left( \real{\langle J_y J_z^* \rangle} \right) 
      - 2\imag{\vec{u}_2\cdot\vec{u}_3^*} \left( \imag{\langle J_y J_z^* \rangle} \right)
      \Bigg\}
\end{align} 
Hence, it should be possible to extract the full 9-element coherence state from a linear inverse-scattering problem with an appropriately PSF kernel:
\begin{align}
    G &= 
    \begin{pmatrix}
        ... & ... & ... & ... & ... & ... \\
        \int_i|\vec{u}_1|^2 & 
        \int_i|\vec{u}_2|^2 &
        \int_i|\vec{u}_3|^2 &
        2 \int_i\real{\vec{u}_1\cdot\vec{u}_2^*} &
        ... &
        -2 \int_i\imag{\vec{u}_2\cdot\vec{u}_3^*} \\
        ... & ... & ... & ... & ... & ...
    \end{pmatrix} 
\end{align}

In fact, an ultra-compact \emph{single-piece} nanophotonic structure should be able to resolve not only polarization states but also extract spatial and spectral information simultaneously from a single measurement. As a proof of principle, we present in Fig.~\ref{fig:supercam} an ``all-in-one super-probe'' which can extract polarization coherence information from up to two spatial points and up to two spectral lines, in which case the target vector $u$ to be reconstructed consists of 36 entries (9 polarization $\times$ 2 spatial $\times$ 2 spectral). The nanophotonic probe has an ultra-compact volume of $10 \times 10 \times 4.5 \lambda^3$, comprising four layers of variable-thickness polymer (refractive index $\approx 1.5$) (Fig.~\ref{fig:supercam}a). A sensor of $10 \times 10$ pixels (pixel area $= \lambda^2$) is located one $\lambda$ away on one side of the probe; on the other side, the two dipoles are positioned one $\lambda$ away from the probe and diagonally separated by $2\sqrt{2}\lambda$ away from each other; the dipoles may emit at $\lambda$, $1.1 \lambda$, or both. The kernel $G$ is a $100 \times 36$ matrix, whose condition number has been optimized to $\kappa_\text{opt} \approx 6.4$ (Fig.~\ref{fig:supercam}b). Here, it is important to emphasize the pivotal role of optimization which greatly improves the noise sensitivity of reconstruction compared to any other non-optimized structure. For example, we found that the kernel $G$ of free space has $\kappa \approx 500$ and that of a correlated random mask (with the correlation length chosen to have similar $\lambda$--2$\lambda$ length-scales to the optimized design) has $\kappa \sim 100$; even modifying the optimized probe by simply discretizing the gray-scale thickness or erasing the thin morphological features may \emph{spoil} $\kappa$ by a factor anywhere between 1.3 and 10.  (That is, any feature-size or binary-thickness constraints must be incorporated during the optimization process using standard techniques~\cite{jensen2011topology}, not imposed after the fact.) We also found that the power captured by the sensor in the presence of the optimized probe is, on average, $\sim 3\times$ greater than for free space or a random mask, indicating that the probe also serves to focus light onto the sensor, further enhancing the noise tolerance. Fig.~\ref{fig:supercam}c shows the reconstruction of two dipoles with two different polarization states and emitting at two wavelengths $\lambda$ (blue) and $1.1\lambda$ (red); the dipoles are circularly polarized along different axes; their coherence matrices are reconstructed under different noise levels $\sigma/\bar{G} = 0.05,~0.20$ with reconstruction errors of $\approx 5\%,~10\%$ respectively.

\section{Summary and outlook}
\label{sec:summary}

The key conclusion of our paper is that optical metastructures designed \emph{in conjunction} with signal processing result in non-obvious light scattering patterns that greatly ease the computational reconstruction.  This results in devices \emph{far more compact} compared to optics-only solutions while being \emph{robust to noise} compared to computation-only designs.   By solving the full (Maxwell) wave equations during the design process, our optimized structure can exploit all available wave physics (non-paraxial scattering, near-field interactions, resonances, dispersion, etc.). We illustrated this idea in the context of examples involving subwavelength imaging and for polarization-state reconstruction, but the same essential ideas can be readily applied to many other systems and computational processing techniques.   In contrast to the many previous metasurface designs that have attempted to mimic and compete with traditional curved lenses~\cite{khorasaninejad2016metalenses}, our scattered fields look nothing like a focal pattern and represent a functionality that is fundamentally distinct from that of conventional optics. Fullwave end-to-end optimization is particularly powerful for problems requiring spectral and polarization information that is discarded by geometric optics, such as polarimetry or hyperspectral imaging.

There are many other sensing/imaging problems that could benefit from this approach.   Our designs in this paper closely resemble \emph{lab-on-a-chip} microscopy.  Related situations arise in ultra-compact opto-fluidic medical sensors, where the probe and sensor must be tightly integrated, the sample is situated only a few wavelengths away from the sensor, and scanning is naturally provided by sample flow~\cite{heng2006optofluidic}.   Inverse design can easily be applied to broad-band problems, and we are especially excited about using it for computational spectroscopy~\cite{yang2019single}, hyper-spectral imaging~\cite{chang2003hyperspectral,kristina2020spectral}, and other broad-band sensing applications.  Our framework can straightforwardly scale to larger 3D freeform structures~\cite{lin2019topology}, accommodate complex high-dimensional objects such as plenoptic light-fields~\cite{ng2005light}, facilitate nonlinear mechanisms such as high dynamic-range imaging~\cite{reinhard2010high}, and generalize to other challenging problems in physics such as nonlinear pulse shaping~\cite{agrawal2001applications} and quantum coherence engineering~\cite{murch2012cavity,bennett20coh}. Optimization can easily incorporate constraints arising from different fabrication processes~\cite{jensen2011topology}.

In this paper, our computational-reconstruction stage consisted of Tikhonov-regularized least-squares fitting, but end-to-end optical design can be coupled with many other computational techniques.    In under-determined systems (many more targets than sensor pixels), a common approach is compressed sensing~\cite{donoho2006compressed} for sparse targets, and techniques for end-to-end optimization with compressed sensing may include differentiable unrolled approximations~\cite{gregor2010learning} or epigraph formulations of basis pursuit denoising~\cite{boyd2004convex}.  One could also employ
deep learning (neural networks) for imaging and other  cognitive tasks (e.g. passive ranging, object recognition); from the perspective of deep learning, the Maxwell solver is simply a specialized ``network stage'' that is differentiable (via adjoint methods) and hence composable with deep-learning software.

Apart from numerical and experimental endeavors, an important theoretical question is to identify the absolute limits to achievable dispersion (spatial or spectral) and condition numbers, given a desired resolution, a design volume $V$, and a dielectric contrast $\Delta \varepsilon$.
Recent approaches for shape-independent bounds to light--matter interactions~\cite{gustafsson2020upper,molesky2020t,kuang2020maximal} may be capable of answering these questions.

\section*{Acknowledgements}
Z.L, C.R.C, R.P, M.S and S.G.J were supported in part by the U.~S.~Army Research Office through the Institute for Soldier Nanotechnologies under award number W911NF-18-2-0048. Z.L and R.P were partially supported by the MIT-IBM Watson AI Laboratory under Challenge 2415.
A.M. was partially supported by a Sloan Fellowship and by the National Science Foundation under award NSF-SNM-1825308.


\end{document}